\documentstyle[12pt]{art12}
\makeatletter
\let\reset@font\empty

\def\indexname{Index}
\def\figurename{Figure}
\def\tablename{Table}
\def\abstractname{Abstract}

\def\@ptsize{0}
\@namedef{ds@11pt}{\def\@ptsize{1}}
\@namedef{ds@12pt}{\def\@ptsize{2}}
\def\ds@twoside{\@twosidetrue
           \@mparswitchtrue}

\def\ds@draft{\overfullrule 5\p@}

\newif\if@titlepage \@titlepagefalse
\def\ds@titlepage{\@titlepagetrue}

\def\ds@twocolumn{\@twocolumntrue}

\newdimen\mathindent
\newif\ifletter
\newif\ifpmb
\newlength{\varind}
\newlength{\figdepth}
\newlength{\figwidth}
\newlength{\secfigwidth}

\newlength{\indentedwidth}

\newcounter{jnl}
\newcounter{yr}
\newcounter{tabtype}
\newcounter{figtype}
\newcounter{eqnval}
\@twosidetrue
\def\ds@draft{\overfullrule 5\p@}
\@options

\def\@normalsize{\@setsize\normalsize{16pt}\xiipt\@xiipt
  \abovedisplayskip 12pt plus3pt minus6pt
  \belowdisplayskip \abovedisplayskip
  \abovedisplayshortskip \z@ plus4pt
  \belowdisplayshortskip 7pt plus4pt minus4pt}
\def\small{\@setsize\small{14pt}\xipt\@xipt
  \abovedisplayskip 10pt plus 3pt minus 4pt
  \belowdisplayskip \abovedisplayskip
  \abovedisplayshortskip \z@ plus3pt
  \belowdisplayshortskip 5pt plus3pt minus 3pt
  \def\@listi{\topsep 5pt plus 3pt minus 3pt\parsep 0pt plus 1pt
         \itemsep \parsep}}
\def\footnotesize{\@setsize\footnotesize{14pt}\xpt\@xpt
  \abovedisplayskip 7pt plus 3pt minus 4pt
  \belowdisplayskip \abovedisplayskip
  \abovedisplayshortskip \z@ plus 2pt
  \belowdisplayshortskip 3pt plus 1pt minus2pt
  \def\@listi{\topsep 4pt plus 2pt minus 2pt\parsep 0pt plus 1pt
         \itemsep \parsep}}
\def\scriptsize{\@setsize\scriptsize{13pt}\ixpt\@ixpt}
\def\tiny{\@setsize\tiny{10pt}\viipt\@viipt}
\def\large{\@setsize\large{18pt}\xivpt\@xivpt}
\def\Large{\@setsize\Large{22pt}\xviipt\@xviipt}
\def\LARGE{\@setsize\LARGE{25pt}\xxpt\@xxpt}
\def\huge{\@setsize\huge{30pt}\xxvpt\@xxvpt}
\def\Huge{\@setsize\Huge{30pt}\xxvpt\@xxvpt}
\normalsize
\skip\footins 12\p@ plus 4\p@ minus 2\p@
\@beginparpenalty -\@lowpenalty
\@endparpenalty -\@lowpenalty
\@itempenalty -\@lowpenalty

\@noskipsecfalse   

\def\section{\@startsection{section}{1}{\z@}{-3.5ex plus -1ex minus
 -.2ex}{2.3ex plus .2ex}{\noindent\reset@font\normalsize\bf\raggedright}}
\def\subsection{\@startsection{subsection}{2}{\z@}{-3.25ex plus -1ex minus
 -.2ex}{1.5ex plus .2ex}{\noindent\reset@font
  \normalsize\it\raggedright\nohyphens}}
\def\subsubsection{\@startsection{subsubsection}{3}{\z@}{-3.25ex plus
-1ex minus -.2ex}{-1em}{\reset@font\normalsize\it\nohyphens}}
\def\paragraph{\@startsection
 {paragraph}{4}{\z@}{3.25ex plus 1ex minus
.2ex}{-1em}{\reset@font\normalsize\it}}
\def\subparagraph{\@startsection
 {subparagraph}{4}{\parindent}{3.25ex plus 1ex minus
 .2ex}{-1em}{\reset@font\normalsize\it}}

\def\@sect#1#2#3#4#5#6[#7]#8{\ifnum #2>\c@secnumdepth
     \let\@svsec\@empty\else
     \refstepcounter{#1}\edef\@svsec{\csname the#1\endcsname.\hskip 1em}\fi
     \@tempskipa #5\relax
      \ifdim \@tempskipa>\z@
        \begingroup #6\relax
          \noindent{\hskip #3\relax\@svsec}{\interlinepenalty \@M #8\par}%
        \endgroup
       \csname #1mark\endcsname{#7}\addcontentsline
         {toc}{#1}{\ifnum #2>\c@secnumdepth \else
                      \protect\numberline{\csname the#1\endcsname}\fi
                    #7}\else
        \def\@svsechd{#6\hskip #3\relax  
                   \@svsec #8\csname #1mark\endcsname
                      {#7}\addcontentsline
                           {toc}{#1}{\ifnum #2>\c@secnumdepth \else
                             \protect\numberline{\csname the#1\endcsname}\fi
                       #7}}\fi
     \@xsect{#5}}
\def\@ssect#1#2#3#4#5{\@tempskipa #3\relax
   \ifdim \@tempskipa>\z@
     \begingroup #4\noindent{\hskip #1}{\interlinepenalty
   \@M #5\par}\endgroup
   \else \def\@svsechd{#4\hskip #1\relax #5}\fi
    \@xsect{#3}}

\def\appendix{\@@par
 \setcounter{section}{0}
 \setcounter{subsection}{0}
 \setcounter{subsubsection}{0}
 \setcounter{equation}{0}
 \setcounter{figure}{0}
 \setcounter{table}{0}
 \def\thesection{Appendix \Alph{section}}
 \def\theequation{\ifnumbysec
      \Alph{section}.\arabic{equation}\else
      \Alph{section}\arabic{equation}\fi}
 \def\thetable{\ifnumbysec
      \Alph{section}\arabic{table}\else
      A\arabic{table}\fi}
 \def\thefigure{\ifnumbysec
      \Alph{section}\arabic{figure}\else
      A\arabic{figure}\fi}}

\leftmargin\leftmargini
\labelwidth\leftmargini\advance\labelwidth-\labelsep
\def\@listI{\leftmargin\leftmargini \parsep 4\p@ plus2\p@ minus\p@
\topsep 8\p@ plus2\p@ minus4\p@
\itemsep 4\p@ plus2\p@ minus\p@}

\let\@listi\@listI
\@listi

\def\@listii{\leftmargin\leftmarginii
 \labelwidth\leftmarginii\advance\labelwidth-\labelsep
 \topsep 3\p@ plus 1\p@ minus 1\p@
 \parsep 0\p@ plus 1\p@
 \itemsep \parsep}
\def\@listiii{\leftmargin\leftmarginiii
 \labelwidth\leftmarginiii\advance\labelwidth-\labelsep
 \topsep 2\p@ plus 1\p@ minus 1\p@
 \parsep \z@ \partopsep 1\p@ plus 0\p@ minus 1\p@
 \itemsep \topsep}
\def\@listiv{\leftmargin\leftmarginiv
 \labelwidth\leftmarginiv\advance\labelwidth-\labelsep}
\def\@listv{\leftmargin\leftmarginv
 \labelwidth\leftmarginv\advance\labelwidth-\labelsep}
\def\@listvi{\leftmargin\leftmarginvi
 \labelwidth\leftmarginvi\advance\labelwidth-\labelsep}

\def\hexnumber@#1{\ifcase#1 0\or 1\or 2\or 3\or 4\or 5\or 6\or 7\or 8\or
 9\or A\or B\or C\or D\or E\or F\fi}
\edef\bffam@{\hexnumber@\bffam}
\mathchardef\bGamma "0\bffam@00
\mathchardef\bDelta "0\bffam@01
\mathchardef\bTheta "0\bffam@02
\mathchardef\bLambda "0\bffam@03
\mathchardef\bXi "0\bffam@04
\mathchardef\bPi "0\bffam@05
\mathchardef\bSigma "0\bffam@06
\mathchardef\bUpsilon "0\bffam@07
\mathchardef\bPhi "0\bffam@08
\mathchardef\bPsi "0\bffam@09
\mathchardef\bOmega "0\bffam@0A

\def\theenumi{\roman{enumi}}

\def\theenumii{\alph{enumii}}
\def\p@enumii{\theenumi.}

\def\theenumiii{\arabic{enumiii}}
\def\p@enumiii{\p@enumii.\theenumii}

\def\p@enumiv{\p@enumiii.\theenumiii}

\def\labelitemi{$\m@th\bullet$}

\def\labelitemiii{$\m@th\ast$}
\def\labelitemiv{$\m@th\cdot$}

\def\verse{\let\\=\@centercr
 \list{}{\itemsep\z@ \itemindent -1.5em\listparindent \itemindent
 \rightmargin\leftmargin\advance\leftmargin 1.5em}\item[]}

\def\quotation{\list{}{\listparindent 1.5em
 \itemindent\listparindent
 \rightmargin\leftmargin\parsep 0\p@ plus 1\p@}\item[]}

\def\descriptionlabel#1{\hspace\labelsep \bf #1}
\def\description{\list{}{\labelwidth\z@ \itemindent-\leftmargin
 \let\makelabel\descriptionlabel}}

\def\enumerate{\ifnum \@enumdepth >3 \@toodeep\else
      \advance\@enumdepth \@ne
      \edef\@enumctr{enum\romannumeral\the\@enumdepth}\list
      {\csname label\@enumctr\endcsname}{\usecounter
        {\@enumctr}\def\makelabel##1{##1\hss}}\fi}
\def\itemize{\ifnum \@itemdepth >3 \@toodeep\else \advance\@itemdepth \@ne
\edef\@itemitem{labelitem\romannumeral\the\@itemdepth}%
\list{\csname\@itemitem\endcsname}{\def\makelabel##1{##1\hss}\topsep=3pt
  \parsep=0pt\listparindent=0pt\itemsep=0pt\partopsep=0pt\rightmargin=0pt
  }\fi}
\tabbingsep \labelsep
\skip\@mpfootins = \skip\footins
\def\titlepage{\@restonecolfalse\if@twocolumn\@restonecoltrue\onecolumn
     \else \newpage \fi \thispagestyle{myheadings}\c@page\z@}

\def\endtitlepage{\if@restonecol\twocolumn \else \newpage \fi}

\newcounter {section}
\newcounter {subsection}[section]
\newcounter {subsubsection}[subsection]
\newcounter {paragraph}[subsubsection]
\newcounter {subparagraph}[paragraph]

\def\thesection {\arabic{section}}

\def\@chapapp{Section}

\def\@pnumwidth{1.55em}
\def\@tocrmarg {2.55em}
\def\@dotsep{4.5}
\def\tableofcontents{\@restonecolfalse\if@twocolumn\@restonecoltrue
 \onecolumn\fi\section*{Contents}{}\thispagestyle{empty}
 \@starttoc{toc}\if@restonecol\twocolumn\fi}
\def\l@section{\@dottedtocline{1}{1.5em}{2.3em}}
\def\l@subsection{\@dottedtocline{2}{3.8em}{3.2em}}
\def\l@subsubsection{\@dottedtocline{3}{7.0em}{4.1em}}
\def\l@paragraph{\@dottedtocline{4}{10em}{5em}}
\def\l@subparagraph{\@dottedtocline{5}{12em}{6em}}
\def\listoffigures{\@restonecolfalse\if@twocolumn\@restonecoltrue\onecolumn
 \fi\section*{List of Figures\@mkboth
 {LIST OF FIGURES}{LIST OF FIGURES}}\@starttoc{lof}\if@restonecol\twocolumn
 \fi}
\def\l@figure{\@dottedtocline{1}{1.5em}{2.3em}}
\def\listoftables{\@restonecolfalse\if@twocolumn\@restonecoltrue\onecolumn
 \fi\section*{List of Tables\@mkboth
 {LIST OF TABLES}{LIST OF TABLES}}\@starttoc{lot}\if@restonecol\twocolumn
 \fi}
\let\l@table\l@figure
\def\@dottedtocline#1#2#3#4#5{\ifnum #1>\c@tocdepth \else
  \vskip \z@ plus .2\p@
  {\leftskip #2\relax \rightskip \@tocrmarg \parfillskip -\rightskip
    \parindent #2\relax\@afterindenttrue
   \interlinepenalty\@M
   \leavevmode
   \@tempdima #3\relax \advance\leftskip \@tempdima
   \hbox{}\hskip -\leftskip
    #4\nobreak\hfill \nobreak \hbox to\@pnumwidth{\hfil
   \rm #5}\@@par}\fi}

\@addtoreset{footnote}{page}
\long\def\@makefntext#1{\parindent 1em\noindent
 \makebox[1em][l]{\footnotesize\rm$\m@th{\fnsymbol{footnote}}$}%
 \footnotesize\rm #1}
\def\@makefnmark{\hbox{${\fnsymbol{footnote}}\m@th$}}
\def\@thefnmark{\fnsymbol{footnote}}
\def\footnote{\@ifnextchar[{\@xfootnote}{\stepcounter{\@mpfn}%
       \begingroup\let\protect\noexpand
       \xdef\@thefnmark{\thempfn}\endgroup
     \@footnotemark\@footnotetext}}
\def\@fnsymbol#1{\ifcase#1\or \dagger\or \ddagger\or \S\or
   \|\or \P\or ^{+}\or ^{\tsty *}\or \sharp
   \or \dagger\dagger \else\@ctrerr\fi\relax}

\def\[{\relax\ifmmode\@badmath\else
 \begin{trivlist}
 \@beginparpenalty\predisplaypenalty
 \@endparpenalty\postdisplaypenalty
 \item[]\leavevmode
 \hbox to\linewidth\bgroup$ \displaystyle
 \hskip\mathindent\bgroup\fi}
\def\]{\relax\ifmmode \egroup $\hfil \egroup \end{trivlist}\else \@badmath \fi}
\def\equation{\@beginparpenalty\predisplaypenalty
 \@endparpenalty\postdisplaypenalty
\refstepcounter{equation}\trivlist \item[]\leavevmode
 \hbox to\linewidth\bgroup $ \displaystyle
\hskip\mathindent}
\def\endequation{$\hfil \displaywidth\linewidth\@eqnnum\egroup \endtrivlist}
\def\eqnarray{\stepcounter{equation}\let\@currentlabel=\theequation
\global\@eqnswtrue
\global\@eqcnt\z@\tabskip\mathindent\let\\=\@eqncr
\abovedisplayskip\topsep\ifvmode\advance\abovedisplayskip\partopsep\fi
\belowdisplayskip\abovedisplayskip
\belowdisplayshortskip\abovedisplayskip
\abovedisplayshortskip\abovedisplayskip
$$\halign to
\linewidth\bgroup\@eqnsel$\displaystyle\tabskip\z@
 {##{}}$&\global\@eqcnt\@ne $\displaystyle{{}##{}}$\hfil    
 &\global\@eqcnt\tw@ $\displaystyle{{}##}$\hfil
 \tabskip\@centering&\llap{##}\tabskip\z@\cr}
\def\endeqnarray{\@@eqncr\egroup
 \global\advance\c@equation\m@ne$$\global\@ignoretrue }
\newcommand{\jl}[1]{\setcounter{jnl}{#1}%
    \ifnum\thejnl=12\global\pmbtrue\fi
    \ifnum\thejnl=15\global\pmbtrue\fi}
\def\journal{\ifnum\thejnl=1 J. Phys.\ A: Math.\ Gen.\
        \else\ifnum\thejnl=2 J. Phys.\ B: At.\ Mol.\ Opt.\ Phys.\
        \else\ifnum\thejnl=3 J. Phys.:\ Condens. Matter\
        \else\ifnum\thejnl=4 J. Phys.\ G: Nucl.\ Part.\ Phys.\
        \else\ifnum\thejnl=5 Inverse Problems\
        \else\ifnum\thejnl=6 Class. Quantum Grav.\
        \else\ifnum\thejnl=7 Network\
        \else\ifnum\thejnl=8 Nonlinearity\
        \else\ifnum\thejnl=9 Quantum Opt.\
        \else\ifnum\thejnl=10 Waves in Random Media\
        \else\ifnum\thejnl=11 Pure Appl. Opt.\
        \else\ifnum\thejnl=12 Phys. Med. Biol.\ %
        \else\ifnum\thejnl=13 Modelling Simul.\ Mater.\ Sci.\ Eng.\
        \else\ifnum\thejnl=14 Plasma Phys. Control. Fusion\
        \else\ifnum\thejnl=15 Physiol. Meas.\
        \else\ifnum\thejnl=16 Sov.\ Lightwave Commun.\
        \else\ifnum\thejnl=17 High Perform.\ Polym.\
        \else\ifnum\thejnl=18 J.\ Hard Mater.\
        \else\ifnum\thejnl=19 J.\ Phys.\ D: Appl.\ Phys.\
        \else\ifnum\thejnl=20 Supercond.\ Sci.\ Technol.\
        \else\ifnum\thejnl=21 Semicond.\ Sci.\ Technol.\
        \else\ifnum\thejnl=22 Nanotechnology\
        \else\ifnum\thejnl=23 Meas.\ Sci.\ Technol.\
        \else\ifnum\thejnl=24 Plasma Source Sci.\ Technol.\
        \else\ifnum\thejnl=25 Smart Mater.\ Struct.\
        \else\ifnum\thejnl=26 J.\ Micromech.\ Microeng.\
        \else\ifnum\thejnl=27 Distrib.\ Syst.\ Engng\
\else Institute of Physics Publishing
\fi\fi\fi\fi\fi\fi\fi\fi\fi\fi\fi\fi\fi\fi\fi
\fi\fi\fi\fi\fi\fi\fi\fi\fi\fi\fi\fi}

\def\catchline{\hfill}

\def\cpyrtline{\hfill}
\def\maketitle{\vspace*{\baselineskip}\vspace{0\p@ plus1fil}
    \noindent Short title: \@shorttitle\par
    \@submitted
    \vspace*{\baselineskip}
    \noindent\today\par\newpage}
\def\@rticle#1#2{\thispagestyle{myheadings}%
     \vspace*{.5pc}%
    {\parindent=\mathindent \bf #1\par}%
     \vspace*{1.5pc}%
    {\exhyphenpenalty=10000\hyphenpenalty=10000
     \Large\raggedright\noindent
     \bf#2\par}\def\@shorttitle{#1}\futurelet\next\sh@rttitle}%
\def\title#1{\def\@shorttitle{#1}%
    \thispagestyle{myheadings}%
    \vspace*{3pc}{\exhyphenpenalty=10000\hyphenpenalty=10000
    \Large\raggedright\noindent
    \bf#1\par}\futurelet\next\sh@rttitle}

\def\article#1#2{\@rticle{#1}{#2}}
\def\review#1{\@rticle{REVIEW \ifpmb\else ARTICLE\fi}{#1}}
\def\topical#1{\@rticle{TOPICAL REVIEW}{#1}}
\def\ireview#1{\@rticle{INTRODUCTORY REVIEW}{#1}}
\def\comment#1{\@rticle{COMMENT}{#1}}
\def\note#1{\@rticle{NOTE}{#1}}
\def\prelim#1{\@rticle{PRELIMINARY COMMUNICATION}{#1}}
\def\letter#1{\@rticle{LETTER TO THE EDITOR}{#1}}
\def\sh@rttitle{\ifx\next[\let\next=\sh@rt
                \else\let\next=\f@ll\fi\next}
\def\sh@rt[#1]{\gdef\@shorttitle{#1}}
\def\f@ll{}
\renewcommand{\author}[1]{\vspace*{1.5pc}%
   \begin{indented}%
   \item[]\normalsize\ifnum\thejnl=8\bf\else\rm\fi\raggedright#1
   \end{indented}%
   \smallskip}
\def\abstract{\vspace{16pt plus3pt minus3pt}
   \begin{indented}
   \item[]{\bf \abstractname.}\quad\rm\ignorespaces}%
\def\endabstract{\end{indented}\vspace{18\p@ plus18\p@}}
\def\submitted{\def\@submitted{\vspace{\baselineskip}%
     \noindent Submitted to: \journal\par}}
\def\@submitted{}
\def\nosections{\vspace{30\p@ plus12\p@ minus12\p@}
    \noindent\ignorespaces}
\newif\ifnumbysec
\def\theequation{\ifnumbysec
      \arabic{section}.\arabic{equation}\else
      \arabic{equation}\fi}
\def\eqnobysec{\numbysectrue\@addtoreset{equation}{section}}
\newenvironment{ceqno}{\begin{ceqno}}{\end{ceqno}}
\def\ceqno{\begin{equation}\begin{array}{@{}*{4}{l}}\dsty}
\def\endceqno{\end{array}\end{equation}}
\def\eqalign#1{\null\vcenter{\def\\{\cr}\openup\jot\m@th
  \ialign{\strut$\displaystyle{##}$\hfil&$\displaystyle{{}##}$\hfil
      \crcr#1\crcr}}\,}
\def\eqalignno#1{\displ@y \tabskip\z@skip
  \halign to\displaywidth{\hspace{5pc}$\@lign\displaystyle{##}$%
    \tabskip\z@skip
    &$\@lign\displaystyle{{}##}$\hfill\tabskip\@centering
    &\llap{$\@lign\hbox{\rm##}$}\tabskip\z@skip\crcr
    #1\crcr}}
\def\cases#1{%
     \left\{\,\vcenter{\def\\{\cr}\normalbaselines\openup1\jot\m@th%
     \ialign{\strut$\displaystyle{##}\hfil$&\tqs
     \rm##\hfil\crcr#1\crcr}}\right.}%
\def\tabular{\def\@halignto{}\@tabular}
\newcommand{\Table}[1]{\def\t@blecap{\caption{#1}}%
   \setcounter{tabtype}{1}\futurelet\next\t@bplace}
\newcommand{\widetable}[1]{\def\t@blecap{\caption{#1}}%
   \setcounter{tabtype}{2}\futurelet\next\t@bplace}
\newcommand{\fulltable}[1]{\def\t@blecap{\caption{#1}}%
   \setcounter{tabtype}{3}\futurelet\next\t@bplace}%
\def\t@bplace{\ifx\next[\let\next=\@tabpl
                 \else\let\next=\@tabnopl\fi\next}
\def\@tabpl[#1]{\begin{table}[#1]\@t@bsize}
\def\@tabnopl{\begin{table}\@t@bsize}
\def\@t@bsize{\ifnum\thetabtype=3\begin{varindent}{0pt}%
   \else\begin{varindent}{\mathindent}\fi
   \t@blecap\lineup\item[]
   \ifnum\thetabtype=1
        \begin{tabular}{@{}l*{15}{l}}
   \else\ifnum\thetabtype=2
        \begin{tabular*}{\indentedwidth}{@{}l*{15}{@{\extracolsep{0pt
plus12pt}}l}}
   \else\begin{tabular*}{\textwidth}{@{}l*{15}{@{\extracolsep{0pt plus12pt}}l}}
   \fi\fi}
\def\endtab{\ifnum\thetabtype=1\end{tabular}
   \else\end{tabular*}\fi\end{varindent}\end{table}}

\newcommand{\centre}[2]{\multicolumn{#1}{c}{#2}}
\newcommand{\crule}[1]{\multispan{#1}{\hrulefill}}
\def\lineup{\def\0{\hbox{\phantom{\footnotesize\rm 0}}}%
    \def\m{\hbox{$\phantom{-}$}}%
    \def\-{\llap{$-$}}}
\long\def\@makecaption#1#2{\vskip 10\p@
 \ifnum\thefigtype=2\begin{varindent}{\@figindent}
 \item[]{\bf #1.} #2
 \end{varindent}\else
 \ifnum\thefigtype=3
 \footnotesize\rm{\bf #1.} #2\else
 \begin{indented}
 \item[]{\bf #1.} #2
 \end{indented}\fi\fi}
\newcommand{\Figure}[1]{\setcounter{figtype}{1}%
    \def\figspace{}\def\figcap{\caption{#1}}%
    \futurelet\next\@figplace}
\def\@figplace{\ifx\next[\let\next=\@figpl
                 \else\let\next=\@fignopl\fi\next}
\def\@figpl[#1]{\begin{figure}[#1]
   \figspace
   \figcap
   \end{figure}}
\def\@fignopl{\begin{figure}
   \figspace
   \figcap
   \end{figure}}

\newcommand{\sidecap}[3]{\setcounter{figtype}{2}%
    \setlength{\figdepth}{#1}\def\@figindent{#2}%
    \def\sidedc@p{\caption{#3}}%
    \futurelet\next\@sidecapplace}
\def\@sidecapplace{\ifx\next[\let\next=\@sidecappl
                 \else\let\next=\@sidecapnopl\fi\next}
\def\@sidecappl[#1]{\begin{figure}[#1]
    \vbox to\figdepth{\vfill
    \sidedc@p}%
    \setcounter{figtype}{1}\end{figure}}
\def\@sidecapnopl{\begin{figure}
    \vbox to\figdepth{\vfill
    \sidedc@p}%
    \setcounter{figtype}{1}\end{figure}}
\newcommand{\side}[3]{\setcounter{figtype}{3}%
    \setlength{\figdepth}{#1}\setlength{\figwidth}{15pc}
    \setlength{\secfigwidth}{15pc}
    \def\firstc@p{\caption{#2}}\def\secondc@p{\caption{#3}}
    \futurelet\next\@sideplace}
\def\@sideplace{\ifx\next[\let\next=\@sidepl
                 \else\let\next=\@sidenopl\fi\next}
\def\@sidepl[#1]{\begin{figure}[#1]
    \vspace*{1.5pc}\vspace*{\figdepth}
    \parbox[t]{\figwidth}{\firstc@p}\hspace*{1pc}%
    \parbox[t]{\secfigwidth}{\secondc@p}
    \setcounter{figtype}{1}\end{figure}}
\def\@sidenopl{\begin{figure}
    \vspace*{1.5pc}\vspace*{\figdepth}
    \parbox[t]{\figwidth}{\firstc@p}\hspace*{1pc}%
    \parbox[t]{\secfigwidth}{\secondc@p}
    \setcounter{figtype}{1}\end{figure}}
\newcommand{\varside}[4]{\setcounter{figtype}{3}%
    \setlength{\figdepth}{#1}\setlength{\figwidth}{#2}%
    \setlength{\secfigwidth}{30pc}
    \addtolength{\secfigwidth}{-\figwidth}
    \def\firstc@p{\caption{#3}}\def\secondc@p{\caption{#4}}
    \futurelet\next\@sideplace}

\newcounter{figure}
\def\thefigure{\@arabic\c@figure}
\def\fps@figure{htbp}
\def\ftype@figure{1}
\def\ext@figure{lof}
\def\fnum@figure{\figurename~\thefigure}
\def\figure{\@float{figure}}
\let\endfigure\end@float
\@namedef{figure*}{\@dblfloat{figure}}
\@namedef{endfigure*}{\end@dblfloat}
\newcounter{table}
\def\thetable{\@arabic\c@table}
\def\fps@table{htbp}
\def\ftype@table{2}
\def\ext@table{lot}
\def\fnum@table{\tablename~\thetable}
\def\table{\@float{table}}
\let\endtable\end@float
\@namedef{table*}{\@dblfloat{table}}
\@namedef{endtable*}{\end@dblfloat}

\def\thebibliography#1{\list
 {\hfil[\arabic{enumi}]}{\topsep=0\p@\parsep=0\p@
 \partopsep=0\p@\itemsep=0\p@
 \labelsep=5\p@\itemindent=-10\p@
 \settowidth\labelwidth{\footnotesize[#1]}%
 \leftmargin\labelwidth
 \advance\leftmargin\labelsep
 \advance\leftmargin -\itemindent
 \usecounter{enumi}}\footnotesize
 \def\newblock{\ }
 \sloppy\clubpenalty4000\widowpenalty4000
 \sfcode`\.=1000\relax}

\def\numrefs#1{\begin{thebibliography}{#1}}
\def\endnumrefs{\end{thebibliography}}

\def\thereferences{\list{}{\topsep=0\p@\parsep=0\p@
 \partopsep=0\p@\itemsep=0\p@\labelsep=0\p@\itemindent=-18\p@
\labelwidth=0\p@\leftmargin=18\p@
}\footnotesize\rm
\def\newblock{\ }
\sloppy\clubpenalty4000\widowpenalty4000
\sfcode`\.=1000\relax
}
\newenvironment{harvard}{\list{}{\topsep=0\p@\parsep=0\p@
\partopsep=0\p@\itemsep=0\p@\labelsep=0\p@\itemindent=-18\p@
\labelwidth=0\p@\leftmargin=18\p@
}\footnotesize\rm
\def\newblock{\ }
\sloppy\clubpenalty4000\widowpenalty4000
\sfcode`\.=1000\relax}{\endlist}
\def\refs{\begin{harvard}}
\def\endrefs{\end{harvard}}
\newenvironment{indented}{\begin{indented}}{\end{indented}}
\newenvironment{varindent}[1]{\begin{varindent}{#1}}{\end{varindent}}
\def\indented{\list{}{\itemsep=0\p@\labelsep=0\p@\itemindent=0\p@
   \labelwidth=0\p@\leftmargin=\mathindent\topsep=0\p@\partopsep=0\p@
   \parsep=0\p@\listparindent=15\p@}\footnotesize\rm}

\def\varindent#1{\setlength{\varind}{#1}%
   \list{}{\itemsep=0\p@\labelsep=0\p@\itemindent=0\p@
   \labelwidth=0\p@\leftmargin=\varind\topsep=0\p@\partopsep=0\p@
   \parsep=0\p@\listparindent=15\p@}\footnotesize\rm}

\def\tabnotes{\ifnum\thetabtype=1\end{tabular}\else\end{tabular*}\fi}
\def\endtabnotes{\end{varindent}\end{table}}

\newif\if@restonecol
\def\theindex{\@restonecoltrue\if@twocolumn\@restonecolfalse\fi
\columnseprule \z@
\columnsep 35\p@\twocolumn[\section*{\indexname}]%
    \@mkboth{{\indexname}}{{\indexname}}%
    \parindent\z@
    \parskip\z@ plus.3\p@\relax\let\item\@idxitem}
\def\@idxitem{\par\hangindent 30\p@}
\def\subitem{\par\hangindent 30\p@ \hspace*{10\p@}}
\def\subsubitem{\par\hangindent 30\p@ \hspace*{20\p@}}
\def\endtheindex{\if@restonecol\onecolumn\else\clearpage\fi}
\def\indexspace{\par \vskip 10\p@ plus 5\p@ minus 3\p@\relax}

\mark{{}{}}

\def\ps@headings{\let\@mkboth\markboth
 \def\@oddfoot{}%
 \def\@evenfoot{}%
 \def\@evenhead{\makebox[\mathindent][l]{\normalsize\rm \thepage}%
  \normalsize\it\rightmark\hfill}%
 \def\@oddhead{\makebox[\mathindent][r]{\hfill}{\normalsize\it\leftmark}\hfill
  \normalsize\rm\thepage}%
}%

\def\ps@myheadings{\let\@mkboth\markboth
 \def\@oddhead{\catchline}%
 \def\@oddfoot{\cpyrtline}%
 \def\@evenhead{}%
 \def\@evenfoot{}%
}

\def\today{\ifcase\month\or
 January\or February\or March\or April\or May\or June\or
 July\or August\or September\or October\or November\or December\fi
 \space\number\day, \number\year}

\def\@begintheorem#1#2{\rm \trivlist \item[\hskip \labelsep{\it #1\ #2.}]}
\def\@opargbegintheorem#1#2#3{\rm \trivlist
      \item[\hskip \labelsep{\it #1\ #2\ (#3).}]}

\def\p@LaTeX{{L\kern-.3em\lower.1em\hbox{$^{\rm A}$}\kern-.15em%
    T\kern-.1667em\lower.7ex\hbox{E}\kern-.125emX}}
\newcommand{\text}[1]{\mbox{#1}}

\newcommand{\nohyphens}{\hyphenpenalty=10000\exhyphenpenalty=10000}

\renewcommand{\qquad}{\hspace*{25pt}}
\newcommand{\tqs}{\hspace*{25pt}}

\def\pt(#1){({\it #1\/})}
\newcommand{\dsty}{\displaystyle}
\newcommand{\tsty}{\textstyle}

\def\;{\protect\psemicolon}
\def\psemicolon{\relax\ifmmode\mskip\thickmuskip\else\kern .3333em\fi}

\newcommand{\opencirc}{\raisebox{2\p@}{\;\circle{5}}}

\newcommand{\fullcirc}{\raisebox{-2\p@}{\Large$\bullet$}}

\newcommand{\boldarrayrulewidth}{1\p@}

\def\bhline{\noalign{\ifnum0=`}\fi\hrule \@height
\boldarrayrulewidth \futurelet \@tempa\@xhline}

\def\@xhline{\ifx\@tempa\hline\vskip \doublerulesep\fi
      \ifnum0=`{\fi}}

\newcommand{\br}{\ms\bhline\ms}
\newcommand{\mr}{\ms\hline\ms}
\newcommand{\ms}{\noalign{\vspace{3\p@ plus2\p@ minus1\p@}}}
\newcommand{\bs}{\noalign{\vspace{6\p@ plus2\p@ minus2\p@}}}
\newcommand{\ns}{\noalign{\vspace{-3\p@ plus-1\p@ minus-1\p@}}}
\newcommand{\es}{\noalign{\vspace{6\p@ plus2\p@ minus2\p@}}\displaystyle}
\newcommand{\JPA}{{\em J. Phys. A: Math. Gen.} }

\ps@headings  \onecolumn
\makeatother
\def\qua{{
        \setlength{\unitlength}{0.1mm}
        \begin{picture}(24,20)(-2,1)
        \put(0,0){\line(1,0){20}}
        \put(0,0){\line(0,1){20}}
        \put(20,20){\line(-1,0){20}}
        \put(20,20){\line(0,-1){20}}
        \end{picture}
        }}
\begin{document}
\jl{1}
\input{epsf.sty}
\begin{flushright}
SISSA Ref. 102/97/EP
\end{flushright}
\vspace*{0.5cm}
\begin{center}
   {\bf \Large 
Spontaneous Breaking of Translational Invariance \\
in One-Dimensional Stationary States\\
on a Ring
 	}\\[15mm]
Peter F. Arndt$\mbox{}^\star$,
Thomas Heinzel$\mbox{}^\star$
and Vladimir Rittenberg$\mbox{}^\star$$\mbox{}^\diamond$
\\[7mm]
$\mbox{}^\star$
Physikalisches Institut, Nu{\ss}allee 12,
53115 Bonn, Germany\\[3mm]
$\mbox{}^\diamond$ SISSA, Via Beirut 2--4, 34014 Trieste, Italy\\[3mm]
{peter@theoa1.physik.uni-bonn.de}\\
{tom@maple.physik.uni-bonn.de}\\
{vladimir@falstaff.he.sissa.it}\\[1.5cm]
\end{center}
\renewcommand{\thefootnote}{\arabic{footnote}}
\addtocounter{footnote}{-1}
\vspace*{2mm}
%
%
%
We consider a model in which positive and negative particles diffuse in 
an asymmetric, $CP$-invariant way on a ring. The positive particles hop
clockwise, the negative counter\-clockwise and 
oppositely-charged adjacent particles may swap positions.
Monte-Carlo simulations and analytic calculations suggest 
that  the model has three phases; a ``pure'' phase in which one has three 
pinned  blocks of only positive, negative particles and vacancies, and in which
translational invariance is spontaneously broken,  a ``mixed'' phase with a  
non-vanishing current in which the three blocks are positive, negative and 
neutral, and a disordered phase without blocks. 
\vspace*{1.5cm}
\begin{flushleft}
cond-mat/9708128 \\
August 1997\\[1cm]
$\mbox{}^\diamond$  work done with partial support of the EC TMR programme,
grant FMRX-CT96-0012
\end{flushleft}
\thispagestyle{empty}
\mbox{}
\newpage
\setcounter{page}{1}
%
\def\figI{
\begin{figure}[tb]
\setlength{\unitlength}{1mm}
\def\setl{\setlength\epsfxsize{10cm}}
\begin{picture}(80,70)
%
\put(20,0){
        \makebox{
                \setl
                \epsfbox{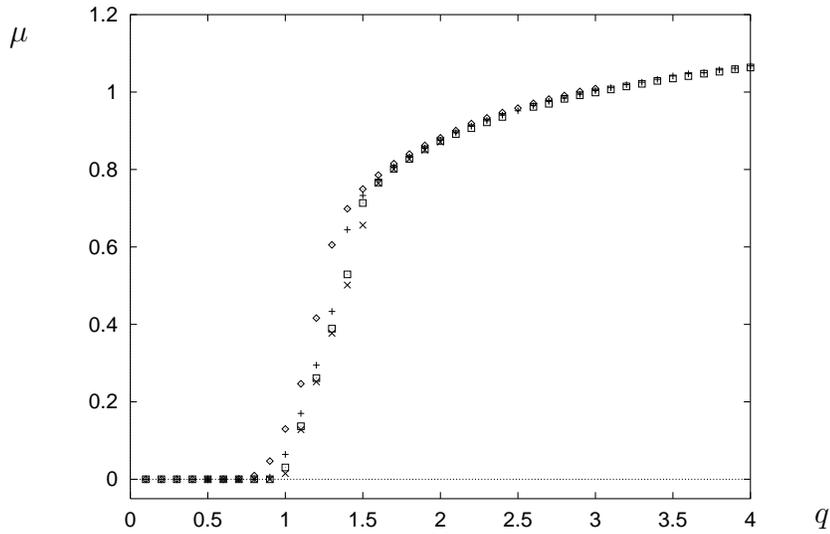}}
        }
\put(20,65){\makebox{$\mu$}}
\put(127,1){\makebox{$q$}}
\end{picture}
\caption{ 
The mobility 
as a function of $q$ for $p=m=0.2$
and $L=100(\diamond)$, $200(+)$, $400(\protect\qua)$, $800(\times)$.
The errors in this and the following figures are less than the size of the symbols.
\label{figI}}
\end{figure}
}
\def\figII{
\begin{figure}[tb]
\setlength{\unitlength}{1mm}
\def\setl{\setlength\epsfxsize{10cm}}
\begin{picture}(80,70)
\put(20,0){
        \makebox{
                \setl
                \epsfbox{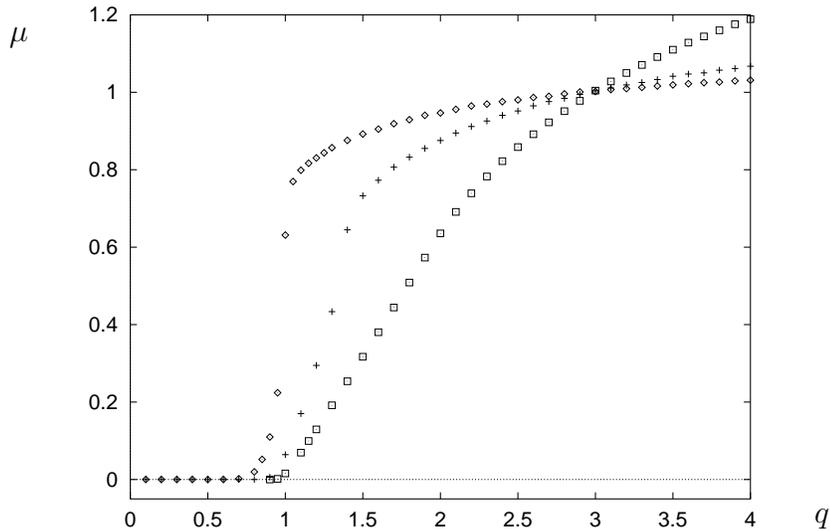}}
        }
\put(20,65){\makebox{$\mu$}}
\put(127,1){\makebox{$q$}}
\end{picture}
\caption{ \label{figII}
The mobility
as a function of $q$ for $p=m=0.1 (\diamond), 0.2 (+) ,0.4 (\protect\qua)$ and $L=200$.
}
\end{figure}
}
%
%
%
\def\figIII{
\setlength{\unitlength}{10pt}
\def\shoi{\begin{picture}(25,10)(-1,-3)
	\put(0,0){\vector(1,0){22.5}}
	\put(0,0){\vector(0,1){8}}
	\put(17,0){\line(0,1){0.3}}
	\put(13,0){\line(0,1){0.3}}
	\put(21,0){\line(0,1){0.3}}
	\put(0,3){\line(1,0){0.3}}
	\put(0,6){\line(1,0){0.3}}
	\put(-0.9,1){\makebox(0,0)[c]{$\scriptstyle b$}}
	\put(-0.9,3){\makebox(0,0)[c]{$\scriptstyle 1-a$}}
	\put(-0.9,6){\makebox(0,0)[c]{$\scriptstyle a$}}
	\put(-2.2,7.1){\makebox(0,0)[c]{\small density}}
	\put(13,-0.7){\makebox(0,0)[c]{$\scriptstyle x$}}
	\put(17,-0.7){\makebox(0,0)[c]{$\scriptstyle \protect\frac{x+1}{2}$}}
	\put(0,-0.7){\makebox(0,0)[c]{$\scriptstyle 0$}}
	\put(21,-0.7){\makebox(0,0)[c]{$\scriptstyle 1$}}
	\put(23,-0.7){\makebox(0,0)[c]{\it z}}
	\thicklines
	\put(13.25,3){\multiput(0,0)(1,0){4}{\line(1,0){0.5}}}
	\put(17.25,6){\multiput(0,0)(1,0){4}{\line(1,0){0.5}}}
	\put(0.25,0.9){\multiput(0,0)(1,0){13}{\line(1,0){0.5}}}
	\put(13,6){\line(1,0){4}}
	\put(17,3){\line(1,0){4}}
	\put(0,1){\line(1,0){13}}
	\thinlines
	\put(6.5,-2){\makebox(0,0)[c]{A}}
	\put(15,-2){\makebox(0,0)[c]{B}}
	\put(19,-2){\makebox(0,0)[c]{C}}
	\end{picture}
	}
\begin{figure}[tb]
\def\sgr{\scriptstyle}
\setlength{\unitlength}{10pt}
\begin{picture}(26,12)(0,0)
\put(10,-1){
        \makebox{\shoi}
	}
\end{picture}
\caption{
\label{figIII}
Density profiles in
the block picture. 
The solid lines denote positive particles. The dashed lines denote
negative particles.
}
\end{figure}
}
\def\figIV{
\begin{figure}[tb]
\setlength{\unitlength}{1mm}
\def\setl{\setlength\epsfxsize{10cm}}
\begin{picture}(80,70)
%
%
\put(20,0){
        \makebox{
                \setl
                \epsfbox{fig4.epsf}}
        }
\put(22,64){\makebox{\large \it c}}
\put(125,-1){\makebox{\large \it z}}
\put(62,38){\makebox{$(a)$}}
\put(62,18){\makebox{$(b)$}}
\end{picture}
\caption{ \label{figIV}
The two-point correlation functions in the ``pure'' (broken) phase
($q=0.5$), with $p=m=0.2$ and
$L=75 (\diamond,\protect\qua)$ and $100 (+,\times)$. 
$(a)$
gives the $c_{0,0}$ and $(b)$ the $c_{+,-}$ correlation function.
}
\end{figure}
}
\def\figVI{
\begin{figure}[tb]
\setlength{\unitlength}{1mm}
\def\setl{\setlength\epsfxsize{10cm}}
\begin{picture}(80,70)
%
%
%
\put(20,0){
        \makebox{
                \setl
                \epsfbox{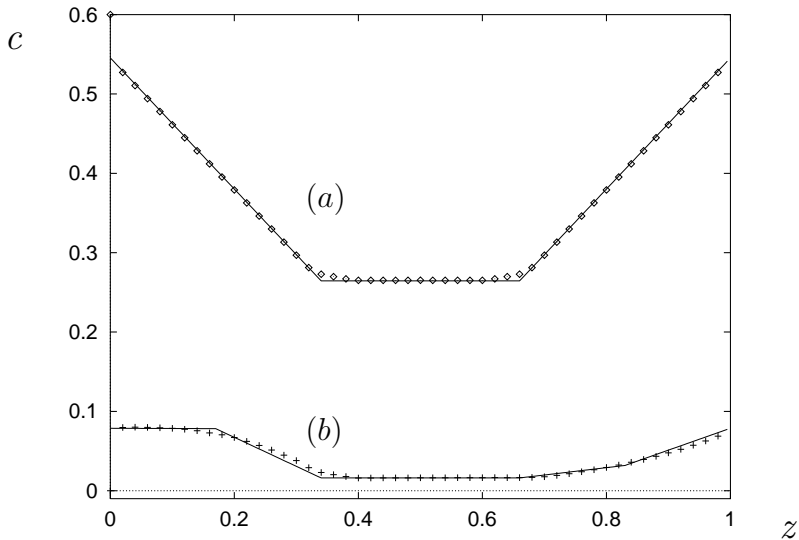}}
        }
\put(22,64){\makebox{\large \it c}}
\put(62,43){\makebox{$(a)$}}
\put(62,12){\makebox{$(b)$}}
\put(125,-1){\makebox{\large \it z}}
\end{picture}
\caption{ \label{figVI}
The two-point functions in the ``mixed'' phase ($q=1.2$) for $p=m=0.2$, $L=600$.
$(a)$
gives the $c_{0,0}$ and $(b)$ the $c_{+,-}$ correlation function.
The solid lines are given by the block picture. The values of $a$
and $x$ are given in Table 1.
}
\end{figure}
}
\def\figVII{
\begin{figure}[tb]
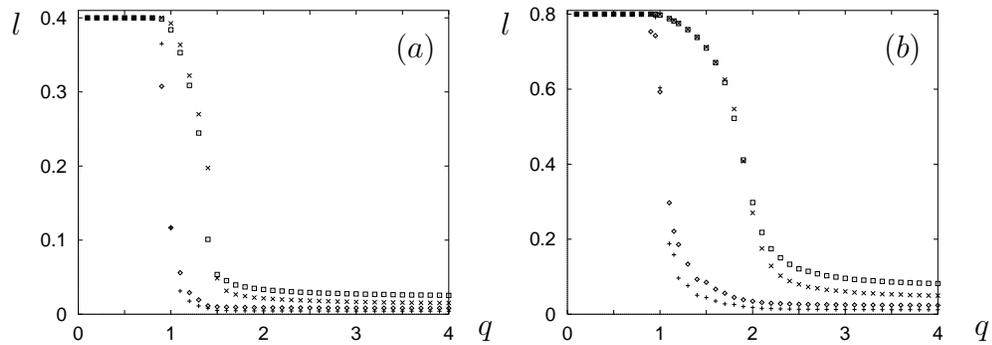

\setlength{\unitlength}{1mm}
\def\setl{\setlength\epsfxsize{6.5cm}}
\begin{picture}(80,42)
%
%
%
%
\put(20,0){
        \makebox{
                \setl
                \epsfbox{fig7a.epsf}}
        }
\put(85,0){
        \makebox{
               \setl
                \epsfbox{fig7b.epsf}}
        }
\put(26,41){\makebox{$l$}}
\put(91,41){\makebox{$l$}}
\put(77,38){\makebox{$(a)$}}
\put(142,38){\makebox{$(b)$}}
\put(88,1){\makebox{$q$}}
\put(154,1){\makebox{$q$}}
\end{picture}
\caption{ \label{figVII}
Average length (in units of $L$) of a string of charged particles ($L=200 (\diamond), 400 (+)$) 
and of the longest string of charged particles ($L=200 (\protect\qua), 400 (\times)$)
as functions of $q$.
$(a)$ has $p=m=0.2$, while $(b)$ has $p=m=0.4$.
}
\end{figure}
}
%
\def\tabi{
\begin{table}[tb]
\caption{Estimates of $a$, $b$ and $x$ for various $1\leq q\leq q_c$.
The numbers were determined by fitting the data for up to $L=1000$.
The errors are of the order 0.03 for $a$ and 
0.01 for $x$. $b$ is given by Eq.(\protect\ref{eqV}).
}
\begin{indented}
\item[]
\begin{tabular}{@{}llllllllll}
\br
\centre{4}{$p=m=0.2$}&&\centre{4}{$p=m=0.4$}
\\
\crule{4}&&\crule{4}
\\
$q$&$a$&$b$&$x$&\qquad&$q$&$a$&$b$&$x$
\\
\mr
1.0 & 1     & 0     &0.60 &&        
1.0 & 1     & 0     &0.20 \\        
1.1 & 0.71  & 0.016 &0.62 &&        
1.2 & 0.67  & 0.04  &0.22 \\        
1.2 & 0.65  & 0.045 &0.66 &&        
1.4 & 0.66  & 0.1   &0.25 \\        
1.3 & 0.63  & 0.077 &0.71 &&        
1.6 & 0.6   & 0.17  &0.30 \\
1.4 & 0.62  & 0.11  &0.76 &&        
1.8 & 0.65  & 0.32  &0.55 \\
\br
\end{tabular}
\end{indented}
\end{table}
}
\noindent
Even if obtained from a master equation with local dynamics, stationary 
states are not necessarily given by Gibbs ensembles. 
In cases where 
one has detailed balance, the hamiltonian may have long range 
interactions. For these reasons, stationary states can exhibit phase 
transitions even in one dimension. Spontaneous $CP$ symmetry breaking was 
seen in open chains, in the two-state partially asymmetric exclusion 
model \cite{DeDoMu,ScDo,DeEvHaPa} (in which $C$ interchanges  particles and vacancies) 
as well as in the three-state model with positive particles, negative particles and 
vacancies \cite{EvFoGoMuii} (in which $C$ interchanges  positive and negative 
particles).
In both models, macroscopic structures in the form of 
shocks exhibiting phase separation appear at the phase transition point 
between the disordered and the broken phases \cite{DeDoMu,Sc,AHRII}. 
In the present letter 
we describe a model defined on a ring which shows spontaneous breaking of 
translational invariance, phase separation, as well as a new type of 
macroscopic structure. We will describe here only some 
highlights of our results, a complete 
presentation being given elsewhere \cite{AHRIV}.

We are considering a ring with $L$ sites numbered $k=0,1,...,L-1$. 
On each site $k$
one may have a positive particle $(+)_k$, a negative particle $(-)_k$ 
or a vacancy $(\,0\,)_k$. 
The time evolution of the system is given by the following rates:
\begin{eqnarray} 
(+)_k\,(\,0\,)_{k+1}\rightarrow (\,0\,)_k\,(+)_{k+1}
\quad\quad&\mbox{rate }1
\nonumber\\ 
(\,0\,)_k\,(-)_{k+1}\rightarrow (-)_k\,(\,0\,)_{k+1}
&\mbox{rate }1
\nonumber\\ 
(-)_k\,(+)_{k+1}\rightarrow (+)_k\,(-)_{k+1}
&\mbox{rate }1
\nonumber\\ 
(+)_k\,(-)_{k+1}\rightarrow (-)_k\,(+)_{k+1}
&\mbox{rate }q
\label{eqI}
\end{eqnarray}
These processes conserve the numbers of positive and negative particles
and are $CP$-invariant. We shall consider only configurations in which the 
average densities of positive and negative particles, denoted by $p$ and $m$ 
respectively, are equal. 
\figI
\figII

We mention that for the model we are considering, the stationary 
state can be computed analytically using the matrix-product approach \cite{DeEvHaPa}
for any value of $q$ \cite{AHRI}. The parameter $q$ determines the physics of 
the problem.

In Fig.\ref{figI} we show, based on results of Monte-Carlo simulations, the 
mobility $\mu$ (defined as the ratio of current to density) as a function of $q$
for $p=m=0.2$ and different values of $L$. 
One can 
distinguish three regions. For $q<1$ the mobility converges exponentially 
to zero. This region will be named, for reasons which will become apparent soon,
as the ``pure'' phase. It is separated by a critical point, $q=1$, from the ``mixed''
phase. At $q=1$, the current vanishes algebraically with $L$ with an
exponent $1.05\pm0.05$. In the ``mixed'' phase, $1<q<1.4$, the mobility is non-zero
and it approaches its limiting value algebraically. The ``mixed'' phase is 
separated by a second critical point $q_c$ from the disordered phase. In  the 
disordered phase, the mobility converges exponentially.
In Fig.\ref{figII} we show the mobility for three densities at $L=200$. One 
notices a remarkable fact. For $q=3$ the mobilities are all equal to 1 
independent of the density or, although not seen in the figure, the 
lattice size. The explanation is simple. Using the matrix-product 
approach which uses representations of a quadratic algebra
\cite{AHRI}, one can show that for $q=3$ one obtains a one-dimensional 
representation of the quadratic algebra. One can
then show that the mobility is equal to 1 for any density and lattice size.
The same approach can be used for $q=4$ where one 
has a two-dimensional representation of the quadratic algebra. 
This allows one to compute the 
mobility and check that the values obtained by Monte-Carlo simulations
are correct. We have also computed the correlation length and 
shown that it is finite. This shows that for $q=4$ one is in the 
disordered phase for any density. 
There are no other finite-dimensional representations of the quadratic algebra
for other values of $q$. However, for other exact results see~\cite{AHRIV}.

Repeating the same simulations for
various lattice sizes we have seen that $q=1$ is the limit of the ``pure''
phase for any density but that the convergence to zero is slower for 
small densities. The second critical point $q_c$ increases with the density.
We found $q_c=1.4\pm 0.1$ for $p=m=0.2$ and $q_c=1.9 \pm 0.1$ for $p=m=0.4$.

We now proceed to clarify the structure of the three phases. We start 
with the ``pure'' phase. 
At $q=0$, a single vacancy is sufficient to break the 
translational invariance of the system. 
At a finite density of 
vacancies, the ground state is infinitely degenerate for $L=\infty$, each configuration
of the kind
\begin{equation}
(\,0\,)\cdots (\,0\,)(+)\cdots (+)(-)\cdots (-)
(\,0\,)\cdots (\,0\,)(+)\cdots (+)(-)\cdots (-)\cdots
\end{equation}
being a stationary state.
At $q$ different from zero and 
finite $L$, if one starts with an arbitrary configuration, the system organizes 
itself into only three blocks
\figIV
\figIII
\tabi
\figVI
\begin{equation}
\label{eqiii}
(\,0\,)(\,0\,)\cdots (\,0\,)(+)(+)\cdots (+)(-)(-)\cdots (-)
\end{equation}
which cover the lattice.
Translational invariance is respected but the blocks hop from one 
position to another with a flip time which increases 
exponentially
with $L$. At the critical point $q=1$ the flip time increases algebraically 
with $L$. This implies that for $q<1$, translational invariance is 
spontaneously broken. Since each of the three blocks contains one kind of 
particle only, we call it the ``pure'' phase.
This situation is illustrated by the behaviour of the 
two-point correlation functions. If we denote by $p(k)$, $m(k)$ and $v(k)$
the concentration of positive particles, 
negative particles and vacancies respectively,
the unconnected correlation functions are
\begin{eqnarray} 
c_{0,0}(k) &=<v(0)\,v(k)\,> 
\nonumber\\
c_{+,-}(k) &= <p(0)m(k)>\,,\quad \mbox{etc.} 
\end{eqnarray} 
In Fig.\ref{figIV} we show the typical behaviour of two correlation functions for 
two small lattice sizes ($L=75$ and $100$). Notice that we have chosen $z=k/L$
rather than $k$
as variable.
It is straightforward to convince oneself that 
these correlation functions correspond to the block picture (\ref{eqiii}) described above. 
The fact that the blocks are already ``pure'' for such small lattice sizes
reflects the exponential behaviour mentioned before.

If $q$ is slightly greater than 1, the charged particles are still 
jammed in the charged blocks (the drift is roughly $q-1$ compared to 1 
in the neutral block and positive particles cannot move in a positive 
block). As a result, we expect the formation of three blocks with a mixed 
composition as shown in Fig.\ref{figIII}. Since we are describing macroscopic blocks 
the appropriate variable is $z=k/L$. For a given average density $p=m$, we have
a neutral block A of length $x$ which is no longer pure but contains
an equal number of positive and negative particles. Next, we have a 
positive block B with a density $a$ of positive particles and $1-a$ of 
negative particles (no vacancies). 
Finally we have a block C, which
is the charge-conjugated version of B.
These blocks are not pinned. 
The beginning of A can be found with equal probability          
in the whole interval $0\leq z<1$. Charge conservation gives
\begin{equation}
\label{eqV}
b=\frac12 (1-\frac{1-p-m}{x})\, .
\end{equation}
In this picture, the mobility can be computed from the densities in the A
block as
\begin{equation}
\label{eqVI}
\mu=\frac{(q-3)b^2+b}{p}\, .
\end{equation}
This picture cannot be exact since a smooth transition of the charged 
density profiles
should exist between the blocks. This is why we have used the A
block and rather than B block to compute the mobility. As $q$ varies between 1 and $q_c$ one 
expects the A block to take over the whole ring.
\figVII

In Fig.\ref{figVI} we show two correlation functions,
obtained from Monte-Carlo simulations and from
the predictions of the block picture, 
for $p=m=0.2$, $q=1.2$ and $L=600$.
The parameters $a$ and $x$ 
were obtained 
by fitting the $c_{+,+}$ correlation function (see Table~1), 
the $c_{0,0}$ and $c_{+,-}$ functions
being then determined. We have checked all 
the correlation functions against the model. The mobility as 
determined by Eq.(\ref{eqVI}) gives 
$\mu=0.209$ as compared to the measured value 
$\mu=0.210$ ($L=600$ sites). The consistency of the block picture with the Monte-Carlo simulation 
was checked for various values of $q$ and densities. Some values for the 
parameters $a$, $b$ and $x$ are given in Table~1. 
We have also observed that for $L\geq 100$ the block picture is already a 
good approximation if one uses $L$-dependent parameters $a$ and $x$ rather than their asymptotic values.

The block picture for the ``mixed'' phase can also be checked in another
way. We have determined the average length $l$ (in units of $L$) of a string of 
charged particles (no vacancies) for various values of $q$. 
The data are
shown in Fig.\ref{figVII} for two densities. 
One notices that for $q<1$ the values 
correspond to the average densities, since we are in the pure phase. 
For 
$q$ slightly greater than 1 the average length of the charged strings 
drops to zero as one expects for the ``mixed'' phase. Next one considers
a different quantity. For each Monte-Carlo configuration one takes the 
length of the longest string of charged particles and averages over configurations.    
This quantity should give an independent estimate for $x$ since the average length of the longest 
string in units of $L$ gives $1-x$. Comparing the values given in Table~1 
with the values in Fig.\ref{figVII} for the corresponding values of $q$, 
one can check
that this is indeed the case.

Finally, we have checked that in the disordered phase, $q>q_c$, no macroscopic
structures are present. This is in agreement
with the exact results obtained for $q=3$ and~4.

A confirmation of the phase structure described above was obtained by studying
the spectrum of the quantum chain hamiltonian \cite{EsRi} describing the 
dynamics of the processes given by Eq.(\ref{eqI}). 
Considering chains of up to 
$10$ sites we numerically investigated the $q$ dependence 
of the first energy levels and observed level crossings 
at values of $q$ compatible with $q_c$.

A complete description of the results presented here can 
be found in \cite{AHRIV}. There, we also give the phase diagram for the case in which 
fewer than three of the four rates in Eq.(\ref{eqI}) are taken as equal.

Before closing this letter, we mention that in a
two-state model on a ring with six-body dynamics, describing a sedimenting 
colloidal crystal, Lahiri and Ramaswamy also found interesting
macroscopic structures \cite{LaRa}.

We would like to thank F.~C.~Alcaraz, M.~den~Nijs, S.~Franz, Y.~Sapir and M.~Virasoro for 
discussions and R.~Behrend for reading the manuscript. 
One of us (V.~R.) would like to thank the EC TMR Programme for 
financial support.
%

%
\end{document}